\pdfoutput=1  
\documentclass[11pt,a4paper]{article}
\usepackage[T1]{fontenc}
\usepackage[utf8]{inputenc}
\usepackage{lmodern}
\usepackage[a4paper,margin=2.5cm]{geometry}
\usepackage[dvipsnames]{xcolor}  

\usepackage{paralist}
\setdefaultenum{1)}{}{}{}
\usepackage[inline]{enumitem}
\usepackage{amsmath}
\usepackage{tablefootnote}
\usepackage{graphicx}
\usepackage{subcaption}
\usepackage{nicematrix}
\usepackage{xargs}
\usepackage{multirow}
\usepackage{booktabs}
\usepackage{stfloats}
\usepackage[many]{tcolorbox}
\usepackage{placeins}

\usepackage[colorinlistoftodos,prependcaption,textsize=tiny]{todonotes}

\makeatletter
\renewcommand
\subsubsection{\@startsection{subsubsection}{3}{\z@}%
{-12\p@ \@plus -4\p@ \@minus -4\p@}%
{-0.5em \@plus -0.22em \@minus -0.1em}%
{\normalfont\normalsize\bfseries\boldmath}}
\makeatother

\newtcolorbox{mybox}[1][]{%
enhanced,
boxsep=0pt,
arc=0.25ex,
colback=white,
boxrule=0.2pt,
leftrule=12pt,
overlay unbroken and first ={%
  \node[rotate=90,
    minimum width=1cm,
    anchor=south,
    font=\footnotesize\sffamily,
    yshift=-14pt,
  white]
  at (frame.west) {Findings};
}
}

\usepackage{xspace}
\newcommand{\nametoolTRACE}{T\textsc{race}\xspace}
\newcommand{\nametoolCLARION}{C\textsc{larion}\xspace}
\newcommand{\nametoolSPADE}{S\textsc{pade}\xspace}
\newcommand{\nametoolSTRACE}{\textsc{strace}}
\newcommand{\nametoolFTRACE}{\textsc{ftrace}\xspace}
\newcommand{\nametoolARTISAN}{A\textsc{rtisan}\xspace}
\newcommand{\nametoolALASTOR}{A\textsc{lastor}\xspace}
\newcommand{\nametoolWinnower}{W\textsc{innower}\xspace}
\newcommand{\nametoolauditd}{\textsc{auditd}\xspace}

\usepackage{soul}
\usepackage{pifont}
\newcommand{\cmark}{\ding{108}}%
\newcommand{\xmark}{\ding{109}}%
\newcommand{\midsym}{\ding{119}}%
\usepackage{makecell} 

\usepackage[numbers,sort&compress]{natbib}
\usepackage{hyperref}

\hypersetup{
pdftitle={A Study of Kernel Telemetry Options for Security-Oriented Provenance},
pdfauthor={Paul R. B. Houssel, Olivier Levillain, Sylvie Laniepce, Nicolas Dejon, Hervé Debar},
pdfkeywords={Provenance, Kernel, Linux, eBPF}
}

\usepackage{authblk}

\begin{document}
\title{A Study of Kernel Telemetry Options for Security-Oriented Provenance}

\author[1,2]{Paul R. B. Houssel}
\author[2]{Olivier Levillain}
\author[1]{Sylvie Laniepce}
\author[1]{Nicolas Dejon}
\author[2]{Hervé Debar}

\affil[1]{Orange Research, France}
\affil[2]{Samovar, Télécom SudParis, Institut Polytechnique de Paris, France}
\affil[ ]{\texttt{\{paul.houssel, s.laniepce, nicolas.dejon\}@orange.com}}
\affil[ ]{\texttt{\{olivier.levillain, herve.debar\}@telecom-sudparis.eu}}

\date{}

\maketitle

\begin{abstract}
Provenance aims to capture the origins, transformations, and interactions of system objects for security and forensic applications. Existing provenance capture approaches still face major challenges and are not yet ready for production environments. In this paper, we first analyze the main kernel telemetry capture approaches, identifying eBPF as the most promising, and complement this analysis with micro benchmarks to assess its performance overhead and the filtering mechanisms used to achieve capture granularity, such as restricting capture to individual containers. Building on this foundation, we then classify, according to the studied capture approaches and filtering methods, eight \textit{provenance systems} and five \textit{capture agents} that could serve as their capture layers, collectively referred to as \textit{tools}. Our study reveals that these \textit{tools} are built on highly heterogeneous capture layers, most of which cannot guarantee the integrity and availability of the captured events, completely failing to meet the requirements of security-oriented use cases.

\end{abstract}

\section{Introduction}
Provenance in computing systems involves recording, analyzing, and leveraging information about the origins, transformations, and interactions of kernel objects, which are accessed and manipulated by user applications through system calls~\cite{carata_primer_2014}. For security-oriented provenance, this information is typically represented as a provenance graph that captures the allocation, modification, and use of these objects, together with the subject responsible for those actions. This historical record is essential for security tasks such as intrusion detection through classification of graph embeddings (e.g., generated by a Graph Neural Network)~\cite{jiang_orthrus_2025}, attack signature identification using graph queries~\cite{blair_automated_2024}, post-mortem forensic auditing~\cite{pasquier_runtime_2018}, and active threat hunting~\cite{li_logkernel_2022}. Compared with identifying malicious signatures in linear audit traces, provenance graphs preserve causal relationships across time and objects, enabling analysts to express and investigate more complex attack behaviors.

\textit{Provenance systems} are designed for whole-host environments, capturing system events at the kernel level to monitor deployed production systems~\cite{perez_systematic_2018}. These approaches still face several challenges~\cite{inam_sok_2023} across their three main architectural layers. At the capture layer, which collects kernel-object allocations and events, these include performance overhead, capture integrity and availability, and the difficulty of determining which kernel objects and system events must be captured to support reliable security analysis. In the reduction layer, incoming telemetry is filtered to achieve a desired level of granularity, such as limiting capture to a specific process or a set of files. At the infrastructure layer, where audit traces are processed and managed, provenance graphs can grow significantly in size, complicating storage and downstream security analysis. These challenges are further amplified in cloud-native environments~\cite{jarkas_container_2025}, where containers share a single kernel and the capture layer must distinguish the events of individual containers from those of the host system to achieve container-level granularity.

In this paper, we first classify the kernel telemetry capture approaches and identify extended Berkeley Packet Filter (eBPF) as the most promising (\S~\ref{sec:capture-approaches-taxonomy}), then empirically evaluate its overhead across program types together with the filtering techniques that enable fine-grained capture (\S~\ref{sec:ebpf-empirical-evaluation}). This qualifies and quantifies how the choice of capture method affects the security applications of provenance graphs, and shows that filtering is essential for production environments, especially in cloud-native computing. We then classify eight \textit{provenance systems} and five \textit{capture agents} that could serve as provenance capture layers, showing that each is built upon the capture approaches and filtering methods studied above, and macro benchmark the open-source subset (\S~\ref{sec:classification}). To our knowledge, this is the first work to systematically study and compare the capture layers across \textit{provenance systems} and their impact on security-oriented provenance.

\section{Related work}
Prior surveys have explored \textit{provenance systems} from several complementary angles. Perez et al.~\cite{perez_systematic_2018} provide a systematic review of provenance challenges across different system layers and environments, offering a generic analysis of these issues. Inam et al.~\cite{inam_sok_2023} extend this effort by formalizing the architectural layers of \textit{provenance systems} and structuring the challenges associated with each layer. Lee et al. investigate provenance in distributed environments~\cite{lee_towards_2015}. Tan et al.~\cite{tan_security_2013} survey integrity mechanisms and encryption schemes for protecting provenance logs in practice. Other studies focus on the detection layer, i.e., the use of provenance graphs for threat detection and forensic analysis. Li et al.~\cite{li_threat_2021} study how to design efficient investigations over large provenance graphs, Zipperle et al.~\cite{zipperle_provenance-based_2022} explore provenance-based intrusion detection systems leveraging machine learning techniques, and Dong et al.~\cite{dong_are_2023} provide an industrial perspective by evaluating the usability and operational effectiveness of such approaches.

These works establish the security value of provenance graphs but largely treat the capture layer as a given, assuming that it records events faithfully and completely. Only the \textit{provenance system} eAudit~\cite{sekar_eaudit_2024} evaluates capture-layer availability, and none assess integrity or completeness. Moreover, although eBPF has become the dominant capture approach and filtering mechanisms are essential in production environments, no prior work empirically compares the performance trade-offs of eBPF program types and filtering strategies for provenance capture. We address these gaps in the following sections.

\section{Kernel telemetry capture approaches}
\label{sec:capture-approaches-taxonomy}
Kernel telemetry captures kernel activity, offering host-wide visibility. Although harder to interpret than application-level telemetry, it is application-agnostic, making it more robust and reliable for security use cases. Vulnerabilities confined to user-space memory (e.g., a heap overflow) may remain invisible in the kernel, but the post-exploitation phase rarely does. From the \textit{provenance systems} and \textit{capture agents} surveyed in \S~\ref{sec:classification}, we identify different capture approaches (Table~\ref{tab:tracing_approach_comparison}) and compare them across five security-relevant properties:
\begin{inparaenum}
\item \emph{portability}: how easily it deploys across kernel versions;
\item \emph{efficiency}: the performance overhead imposed;
\item \emph{visibility}: the coverage of telemetry data;
\item \emph{safety}: the risk of system instability (e.g., kernel panics, memory corruption); and
\item \emph{robustness}: whether it introduces attack vectors affecting trace integrity and availability.
\end{inparaenum}
Because the \textit{robustness} challenges largely generalize across capture approaches, we examine this property across all of them in detail (\S~\ref{sec:robustness}). We distinguish kernel telemetry by its capture location: from the user space or kernel space.

\begin{table}[h!]
\caption{Capture approaches for kernel telemetry (\cmark{} good, \midsym{} moderate, \xmark{} poor).}
\label{tab:tracing_approach_comparison}
\centering
\resizebox{0.8\linewidth}{!}{
  \begin{tabular}{ccc@{\hspace{1em}}c@{\hspace{1em}}c@{\hspace{1em}}c@{\hspace{1em}}c@{\hspace{1em}}c}
    \toprule
    \textbf{location}&\textbf{approach}&\textbf{\textit{portability}}&\textbf{\textit{efficiency}}&\textbf{\textit{visibility}}&\textbf{\textit{safety}}&\textbf{\textit{robustness}}&\textbf{example}\\
    \midrule
    \multirow{2}{*}{\makecell{user space}}&\makecell{\texttt{ptrace}}&\midsym&\xmark&\xmark&\midsym&\xmark&\nametoolALASTOR~\cite{datta_alastor_2022}\\
    \cmidrule{2-8}
    &fs snapshot&\midsym&\cmark&\xmark&\cmark&\xmark&\nametoolARTISAN~\cite{yu_cost-effective_2024}\\
    \midrule
    \multirow{4}{*}{\makecell{kernel\\space}}&\makecell{integrated systems}&\xmark&\midsym&\midsym&\cmark&\midsym&\nametoolCLARION~\cite{chen_clarion_2021}\\
    \cmidrule{2-8}
    &out-of-tree kernel module&\xmark&\cmark&\cmark&\midsym&\midsym&LTTng~\cite{noauthor_lttng_2005}\\
    \cmidrule{2-8}&eBPF&\midsym&\cmark&\cmark&\cmark&\midsym&ProvBPF~\cite{lim_secure_2021}\\
    \bottomrule
\end{tabular}}
\end{table}

\subsection{Capture from user space}
Kernel telemetry can be collected from user space using two main approaches. The first uses the \texttt{ptrace} system call to attach to a process and monitor its system calls and signals. Context switching between user and kernel space renders it \textit{inefficient}, and it lacks \textit{visibility} into kernel internals, threads, and lower-level events such as network packets. In restricted environments without host access, such as serverless containers, \texttt{ptrace} remains the only capture mechanism; e.g., \nametoolALASTOR~\cite{datta_alastor_2022} employs \nametoolSTRACE~\cite{noauthor_strace_1991} (which relies on \texttt{ptrace}). Although long available in Linux, its \textit{portability} is only partial, as the system-call ABI evolves~\cite{houssel_towards_2025}. Prior privilege-escalation vulnerabilities highlight inherent \textit{safety} risks~\cite{connor_pku_2020}.

The second approach infers telemetry from user-space observations; e.g., \nametoolARTISAN~\cite{yu_cost-effective_2024} captures periodic file system snapshots to infer causality from the differences. By avoiding continuous interception, it is more \textit{efficient} than \texttt{ptrace}-based methods and introduces no \textit{safety} issues. However, its \textit{visibility} is limited to file-system state changes, omitting other kernel events and objects (e.g., process and network activity); dependence on specific OS or kernel versions limits \textit{portability}; and \textit{robustness} is even weaker, as an attacker can modify the file system between snapshots, leaving no trace in the inferred telemetry.

\subsection{Capture from kernel space}
\label{sec:fromthekernel}
Kernel telemetry can also be collected directly from within the kernel, through integrated systems, out-of-tree kernel modules, or eBPF programs. Operating in the kernel generally provides \textit{efficiency} and deep \textit{visibility} into kernel telemetry.

\subsubsection{Integrated capture systems}
\nametoolFTRACE and \nametoolauditd are \textit{tools} built into the kernel source. \nametoolauditd serves as the capture layer for \textit{provenance systems} such as \nametoolWinnower~\cite{hassan_towards_2018}, \nametoolCLARION~\cite{chen_clarion_2021}, and \nametoolTRACE~\cite{irshad_trace_2021}. Despite adoption, these approaches fall short for three reasons:
\begin{inparaenum}
\item kernel source modifications require lengthy peer-review, ensuring \textit{safety} but delaying updates and hindering \textit{portability};
\item this \textit{portability} limit forces adaptations into user space, restricting fine-grained capture to the reduction layer, which we later show is less efficient than filtering in the capture layer; and
\item originally built for debugging or targeted audits rather than continuous whole-system capture, they are less \textit{efficient} under the sustained event rates of long-running, system-wide security monitoring~\cite{sekar_eaudit_2024}.
\end{inparaenum}

The kernel also hosts Linux Security Modules (LSMs) e.g., SELinux~\cite{smalley_implementing_2001}, which enforce Mandatory Access Control (MAC) via dedicated \textit{LSM} interfaces. While LSMs are typically in-tree modules shipped within the kernel source, they can also be distributed as out-of-tree kernel modules or eBPF programs.

\subsubsection{Out-of-tree kernel modules}
Out-of-tree kernel modules are compiled outside the main kernel source tree and mapped into the kernel's address space, extending functionality without modifying the source code. While most act as drivers, some implement security features~\cite{angelakopoulos_firmsolo_2023}. They are the least \textit{portable} approach considered: a module's implementation is tightly bound to a single kernel version, depending on internal kernel APIs and data structures that change across releases~\cite{houssel_towards_2025}, so every target kernel requires a separately adapted build. \textit{Safety} concerns also arise: unlike in-tree modules, they bypass the kernel's peer-review process, and a poorly implemented module can cause kernel panics or memory corruption (e.g., out-of-bounds writes).

\subsubsection{eBPF programs}
eBPF programs are a \textit{safer} alternative as they extend the kernel through a non-Turing-complete language that guarantees termination (bounded loops, limited instruction counts)~\cite{vishwanathan_verifying_2023}, while typing and fixed-size allocations enforce memory safety. eBPF generalizes the classic Berkeley Packet Filter from network packet inspection to arbitrary kernel instrumentation. It is partly \textit{portable}: available since kernel~v3.18, though the relevant program types arrived later (from v4.1; Table~\ref{tab:ebpf_program_type_comp}) and their attach interfaces may change across versions, constraining version-agnostic deployment~\cite{houssel_towards_2025,zhong2025revealing}. We identify four major program types for provenance capture (\textit{LSM}, \textit{tracepoint}, \textit{kprobe} and \textit{tracing}), each attaching to specific kernel interfaces.

\textit{LSM} programs attach to \textit{LSM} interfaces, historically used by LSMs such as SELinux to enforce MAC policies, where multiple LSMs can be chained, each invoked sequentially to allow or deny the operation. eBPF \textit{LSM} programs use the \texttt{LSM\_MAC} attach type (all processes) or \texttt{LSM\_CGROUP} (a given \textit{cgroup}), dispatched through a BPF \textit{trampoline}, a stub the kernel patches in at attach time to jump directly into the program. These interfaces offer more stable ABIs than kernel functions~\cite{houssel_towards_2025}, but a single MAC operation may map to multiple system calls, preventing system-call granularity.

\textit{tracepoint} programs attach to predefined, stable \textit{tracepoint} interfaces that cover system call entry and exit points as well as other kernel semantic events, via a static \texttt{NOP} patched into a call upon attachment. The patched call does not reach the eBPF program directly: it enters the kernel's generic tracepoint dispatcher, which on each firing copies the site's fixed arguments into a record and passes it to every registered handler.

\textit{kprobe} programs attach to nearly any kernel function entrance (\texttt{kprobe}) or exit (\texttt{kretprobe}), as well as to user-space function entry (\texttt{uprobe}) and exit (\texttt{uretprobe}). The target instruction is replaced with an interrupt (e.g., \texttt{INT3} on x86), which triggers the eBPF program before resuming execution. While offering extensive \textit{visibility}, including over \textit{LSM} hooks (functions with the ``\texttt{security\_}'' prefix), they incur additional overhead from CPU interruptions and attach to less stable interfaces that may change across versions.

\textit{tracing} programs cover a similar scope to \textit{kprobe} (excluding user-space functions) but, like \textit{LSM} programs, attach through a BPF trampoline rather than an interrupt: with \texttt{CONFIG\_FUNCTION\_TRACER}, attachment turns the entry \texttt{NOP} into a trampoline call that runs the \texttt{fentry}/\texttt{fexit} programs around the function, avoiding \textit{kprobe}'s CPU interruptions. Like \textit{tracepoint}, it starts from a patched \texttt{NOP}, but the target differs: instead of the generic dispatcher with its per-firing argument copying, the trampoline is generated at attach time for that specific program and function, handing the function's register-held arguments straight to the program and leaving minimal per-event work.

\begin{table}[h!]
\caption{Comparison of eBPF program types used for provenance capture layers.}
\label{tab:ebpf_program_type_comp}
\centering
\resizebox{0.9\linewidth}{!}{
  \begin{tabular}{cccccccccc}
    \toprule
    \makecell{\textbf{Program}\\\textbf{type}}&\makecell{\textbf{Attach}\\\textbf{type}}&\makecell{\textbf{TOCTOU}\\\textbf{resistance}}&\multicolumn{5}{c}{\textbf{Granularity}}&\textbf{Stability}&\makecell{\textbf{Kernel}\\\textbf{version \(\geq\)}}\\
    &&&\makecell{system\\calls}&\makecell{kernel\\ function}&\makecell{user-space\\function}&\makecell{MAC\\operation}&\textit{cgroup}&&\\
    \midrule
    \multirow{2}{*}{\textit{LSM}}&\texttt{LSM\_MAC}&\cmark&\xmark&\xmark&\xmark&\cmark&\xmark&\cmark&5.7\\
    &\texttt{LSM\_CGROUP}&\cmark&\xmark&\xmark&\xmark&\cmark&\cmark&\cmark&6.0\\
    \midrule
    \textit{tracepoint}&\texttt{tracepoint}&\midsym&\cmark&\midsym&\xmark&\xmark&\xmark&\cmark&4.7\\
    \midrule
    \multirow{2}{*}{\textit{kprobe}}&\texttt{kprobe/kretprobe}&\midsym&\cmark&\cmark&\xmark&\cmark&\xmark&\xmark&4.1\\
    &\texttt{uprobe/uretprobe}&\xmark&\xmark&\xmark&\cmark&\xmark&\xmark&\xmark&4.3\\
    \midrule
    \textit{tracing}&\texttt{fentry/fexit}&\midsym&\cmark&\cmark&\xmark&\cmark&\xmark&\xmark&5.5\\
    \bottomrule
\end{tabular}}
\end{table}

Table~\ref{tab:ebpf_program_type_comp} summarizes eBPF program types. Among them, \textit{LSM} programs stand out as the most suitable choice for kernel-telemetry provenance: they enforce policies at a granular level, particularly within containers through \textit{cgroup} attachments, and are more stable than \textit{kprobe} and \textit{tracing} programs, as they adapt to internal kernel changes~\cite{houssel_towards_2025}. Their scope, however, is limited to MAC operations, raising the question of whether this is sufficient to construct complete provenance graphs. The Linux kernel v6.0 requirement for \textit{cgroup}-attached \textit{LSM} programs is not a significant barrier: among kernels older than 6.0, only 5.15 and 5.10 still receive long-term support, and both reach end of life in December~2026~\cite{kernel_org_releases}.

\subsection{A closer look at the \textit{robustness} property}
\label{sec:robustness}
Among user-space methods, \textit{robustness} is weak. With \texttt{ptrace}-based tracing, a malicious process can occupy its single tracer slot (e.g., with \texttt{PTRACE\_TRACEME}),  blocking other tracers. With snapshot-based approaches, an attacker may execute critical steps between two snapshots, leaving no evidence in the inferred telemetry. Furthermore, user-space methods can be exposed to unprivileged processes~\cite{connor_pku_2020}, whereas kernel-based approaches run behind strict privilege boundaries and are therefore more \textit{robust}. Kernel modules and eBPF programs further strengthen \textit{robustness}, as by default only privileged users can load them.

However, all approaches that trace system calls are vulnerable to Time-Of-Check to Time-Of-Use (TOCTOU) race-condition attacks~\cite{guo_trace_2022}: an adversary can alter system call arguments after the check but before kernel processing, compromising capture integrity (e.g., opening a suspicious file while a benign filename is traced). The race window can be widened, for instance via a custom page-fault handler (\texttt{userfaultfd}) or a delayed TCP handshake on a controlled server. This affects any mechanism hooking system call entry points that accept user-space pointers. In Linux~6.14 (Table~\ref{tab:pointer_prevalence}), 69\% of system calls take at least one such pointer and are therefore vulnerable; among them, 96 point to a filename or file handle and 8 to a socket address, enough to conceal major attack steps. The rest are mostly buffers or call-specific option structures and flags, which security telemetry rarely captures, further limiting visibility. The issue is even more pronounced in security monitoring \textit{tools}: over 70\% of system calls in Falco and Tracee attack signatures accept a user-space pointer (Appendix~\ref{sec:appendix_robustness}).

\begin{table}[h!]
\centering
\caption{Prevalence of system calls with user-space pointer arguments.}
\label{tab:pointer_prevalence}
\resizebox{\linewidth}{!}{
  \begin{tabular}{lc@{\hspace{1.2em}}c@{\hspace{1.2em}}c}
    \toprule
    & \# system calls & \multicolumn{2}{c}{\# of those with a pointer that is} \\
    \cmidrule(lr){2-2}\cmidrule(lr){3-4}
    Set of system calls & with $\geq$1 user-space pointer & file-related & socket-address \\
    \midrule
    All system calls (Linux v6.14) & 282/407 (69\%) & 96 & 8 \\
    $\in$ Falco attack signatures~\cite{Falco}  & 30/41 (73\%) & 20 & 5 \\
    $\in$ Tracee attack signatures~\cite{noauthor_tracee_2020} & 14/19 (74\%) & 9  & 1 \\
    \bottomrule
\end{tabular}}
\end{table}

eBPF \textit{LSM} programs distinguish themselves here: security checks occur after arguments have been copied into kernel space, at deeper points along the execution path, conferring stronger resistance to TOCTOU races~\cite{guo_trace_2022,watson_exploiting_2007}. The three other eBPF program types remain vulnerable when attached to system calls.

\section{Empirical evaluation of eBPF capture}
\label{sec:ebpf-empirical-evaluation}
Based on our evaluation of capture approaches (Table~\ref{tab:tracing_approach_comparison}), eBPF stands out as the most promising for security-oriented provenance: it is the only approach that fully satisfies \textit{efficiency}, \textit{visibility}, and \textit{safety}, while its \textit{robustness} and \textit{portability} depend on which kernel interfaces are instrumented. We therefore empirically evaluate its overhead across four eBPF program types: \textit{tracepoint}, \textit{kprobe}, \textit{tracing} and \textit{LSM} (\S~\ref{sec:overhead-program-type}) and across the filtering methods that achieve capture granularity, an important and understudied aspect for production environments (\S~\ref{sec:capture_granularity}).

\subsection{Experimental setup}
\label{sec:experimental-setup}
All experiments are conducted on a 24-core host running Debian~12 (Linux~6.1.0-amd64). We test each capture configuration with three benchmarks representing common production workloads:
\begin{inparaenum}
\item \textit{network} workload, using \textit{httperf}~\cite{mosberger_httperftool_1998} to generate 50k HTTP connections from a client directly connected to a local python HTTP server;
\item \textit{file} workload, using \textit{postmark}~\cite{katcher_postmark_1997} to write 50k files of 512~KB each, 500 accessed simultaneously; and
\item \textit{process} workload, using \textit{shbm}~\cite{sekar_eaudit_2024} to execute the \texttt{echo} binary sequentially 50k times.
\end{inparaenum} The traced services run inside Docker containers. We quantify performance overhead as the relative CPU-time increase between a traced workload and an untraced baseline, averaged over 10~runs after 10 warm-up iterations, with fixed P-states and disabled C-states. All programs share the same eBPF maps, layouts, and user-space logger, keeping memory overhead constant, and each records the PID, UID, TGID, executable name, and \textit{cgroup} ID of the triggering task, ensuring semantic equivalence.

\subsection{Overhead by eBPF program type}
\label{sec:overhead-program-type}
Because the program types cannot all attach to the same hooks (\textit{LSM} programs cannot attach to system calls, and \textit{tracepoint} programs cannot attach to \textit{LSM} hooks), we compare them at two locations. At the \textit{system call location}, we instrument \textit{tracing}, \textit{kprobe}, and \textit{tracepoint} programs; at the \textit{LSM location} we attach \textit{LSM} programs to the \textit{LSM} interfaces and \textit{kprobe} and \textit{tracing} programs to the kernel functions that invoke them (prefixed with \texttt{security\_}). Some programs embed filters so that all capture exactly the same semantic event (connection acceptance for \textit{network}, file creation for \textit{file}, and task cloning for \textit{process}) and our eBPF \textit{LSM} program is registered last in the LSM chain so that it only captures operations allowed by all preceding LSMs. For every program type but \textit{LSM}, capturing these events requires a dual attachment, at entry and exit, to respectively read the arguments and discard rejected operations: system calls rejected by the kernel or by an LSM enforcing MAC~\cite{craun2025pairwise}, and, at the \textit{LSM location}, operations denied by another LSM such as SELinux after the \texttt{security\_} wrapper has been entered. We account for this added cost in our measurements. Table~\ref{tab:traced_interfaces} summarizes the kernel interfaces instrumented in this experimental setup.

\begin{table}[h!]
\centering
\caption{Kernel interfaces instrumented in our experimental setup.}
\label{tab:traced_interfaces}
\resizebox{\linewidth}{!}{
  \begin{tabular}{cc@{\hspace{2.5em}}c@{\hspace{2.5em}}c@{\hspace{1em}}c}
    \toprule
    \multirow{2}{*}{\makecell{Tracing\\location}}&\multicolumn{3}{c}{Traced interface per workload (captured event)}&\multirow{2}{*}{\makecell{eBPF program\\type}}\\
    \cmidrule{2-4}
    &\makecell{\textit{network}\\(accept conn.)}&\makecell{\textit{file}\\(file creation)}&\makecell{\textit{process}\\(clone)}&\\
    \midrule
    \textit{system call}&\makecell{\texttt{accept},\\\texttt{accept4}}&\makecell{\texttt{creat}, \texttt{open}\textsuperscript{a},\\\texttt{openat}\textsuperscript{a}, \texttt{openat2}\textsuperscript{a}}&\makecell{\texttt{clone},\\\texttt{clone3}}&\makecell{\textit{tracepoint},\\\textit{kprobe}, \textit{tracing}}\\
    \midrule
    \multirow{2}{*}{\textit{LSM}}&\multirow{2}{*}{\texttt{socket\_accept}}&\multirow{2}{*}{\texttt{inode\_init\_security}}&\multirow{2}{*}{\texttt{task\_alloc}\textsuperscript{b}}&\textit{LSM}\\
    &&&&\textit{kprobe}, \textit{tracing}\\
    \bottomrule
  \end{tabular}
}
\smallskip
{\scriptsize \textsuperscript{a}\,\texttt{O\_CREAT} flag filtered to capture only file creations. \textsuperscript{b}\,\texttt{PF\_KTHREAD} flag filtered to sieve kernel threads.}
\end{table}

At the \textit{system call location} with entry-only attachment, \textit{tracing} is the most efficient across all workloads: 0.76\% (\textit{network}), 8.12\% (\textit{file}), and 1.15\% (\textit{process}), against 1.43\%, 8.63\%, and 1.22\% for \textit{tracepoint} and 1.66\%, 9.10\%, and 1.30\% for \textit{kprobe} programs. This ordering follows the attachment mechanism: \textit{tracing} and \textit{tracepoint} rely on \texttt{NOP} patching rather than \textit{kprobe}'s CPU interruptions, \textit{tracing} stays ahead because its trampoline calls the program directly from the function prologue, whereas \textit{tracepoint}'s generic dispatcher copies a fixed argument record at every firing, an extra per-event cost.

\begin{figure}[h!]
\centering
\includegraphics[width=0.8\linewidth]{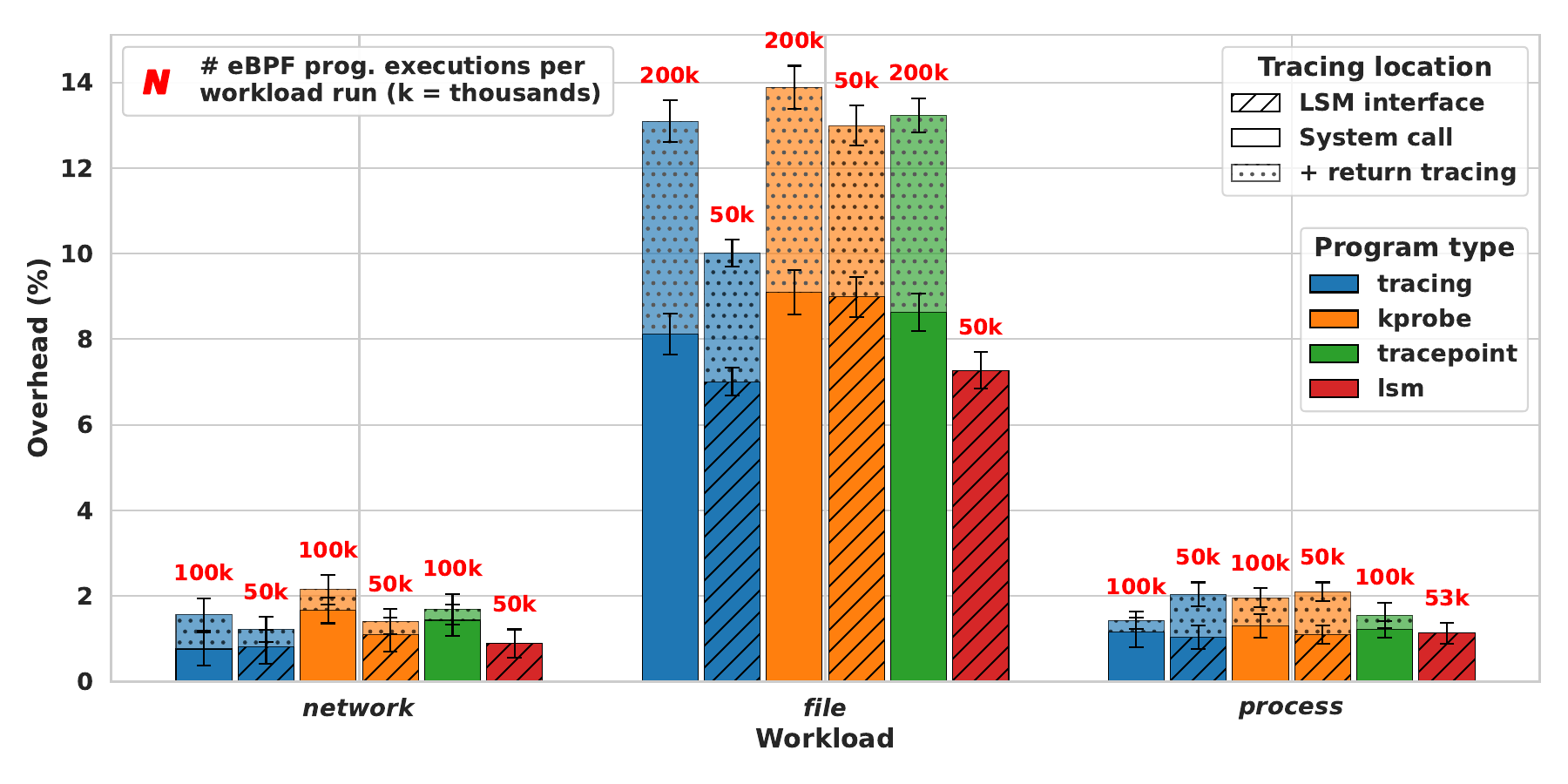}
\caption{Performance overhead among eBPF programs.}
\label{fig:experiment_ebpf_program_type}
\end{figure}

Fig.~\ref{fig:experiment_ebpf_program_type} reports these results. The solid (\textit{system call location}) and hatched (\textit{LSM location}) bars give the cost of the entry attachment, which captures the arguments; the dotted segments give the additional cost of the exit attachment, needed to observe the verdict and discard rejected events. Only \textit{LSM} programs obtain both with a single attachment: registered last in the chain, they execute only for operations allowed by all other LSMs.

At the \textit{LSM location} with entry-only attachment, \textit{tracing} remains the most efficient across all workloads: 0.82\% (\textit{network}), 7.01\% (\textit{file}), and 1.04\% (\textit{process}), against 1.10\%, 8.99\%, and 1.10\% for \textit{kprobe} and 0.89\%, 7.27\%, and 1.13\% for \textit{LSM} programs, although the gap with \textit{LSM} is negligible given the variance and explained by their attachments: \textit{tracing} programs attach a trampoline directly to the \texttt{security\_} function prologue, whereas \textit{LSM} programs, attached through the same mechanism, are additionally dispatched through the LSM framework. This advantage does not survive verdict capture: also tracing the \texttt{security\_} wrapper's return raises \textit{tracing} to 1.22\%, 10.01\%, and 2.04\% and \textit{kprobe} to 1.40\%, 12.99\%, and 2.10\%, making \textit{LSM} programs the cheapest at this location across all workloads. The \textit{LSM location} is already cheaper than the \textit{system call location} even before counting the second attachment, most markedly for the \textit{file} workload: the \textit{LSM} interface \texttt{inode\_init\_security} fires only on file creation (50k executions), whereas system-call programs must inspect every \texttt{open} (100k, as each file is also reopened for reading) for the \texttt{O\_CREAT} flag to isolate the 50k creations, a count doubled to 200k by the entry/exit dual attachment.

For provenance, this second attachment is not optional: an operation recorded at entry may still be denied by the kernel or by an LSM such as SELinux. These blocked attempts are valuable to observability, revealing what an attacker tried, but a provenance graph must contain only the causal events that actually occurred. In sum, once the verdict is captured (dotted segments), \textit{tracing}, \textit{kprobe}, and \textit{tracepoint} programs roughly double their per-event work~\cite{craun2025pairwise}, and \textit{LSM} programs, despite \textit{tracing}'s lower per-hook cost, are the most efficient for provenance capture, obtaining the same semantic event (arguments and verdict) with a single attachment.

Overall, performance overhead is highest in the \textit{file} workload, significantly exceeding that of the \textit{network} and \textit{process} workloads. This is due not only to the heavier argument parsing but also to the inherent latency of kernel operations like \texttt{inode\_init\_security}, which involve the label recomputation and initialization of the new inode~\cite{yamauchi_lsmpmon_2012,zhang_analyzing_2021}.


\subsection{Overhead by filtering granularity}
\label{sec:capture_granularity}
In this section, we study container granularity. While our scope is not limited to cloud-native computing, containers are the prevalent deployment unit in production and make the need for granularity concrete: the capture layer must separate a monitored workload's events from those of the host and co-located workloads, e.g., to trace a single exposed service on a multi-tenant host. The filtering methods we compare apply unchanged to other granularities.

Provenance capture can operate at different granularities, targeting events (e.g., specific system calls), objects (e.g., specific processes or users), or containers. Such granularity is essential in production: capturing all kernel activity inflates overhead and data volume, whereas filtering at the capture layer retains only the subset relevant to the security analysis. eBPF programs achieve event granularity by attaching to specific interfaces, such as system calls, kernel or user-space functions, and to MAC operations (Table~\ref{tab:ebpf_program_type_comp}), and object granularity by filtering event attributes within the program itself. \textit{LSM} programs can additionally achieve container granularity by attaching to a specific \textit{cgroup} (unlike other eBPF program types): every container is assigned a dedicated \textit{cgroup} by the OCI-standard runtimes (e.g., \texttt{runc}) underlying Docker, Podman, containerd, and CRI-O. Its file descriptor, located in the \textit{cgroup} file system (\texttt{/sys/fs/cgroup}), can be used to create an eBPF link~\cite{cgroupattachementbpfgitcommit} that attaches an eBPF program to the \textit{cgroup}, so that it triggers only for events from processes within that \textit{cgroup}.

\textit{Cgroup} attachment filters events before the eBPF program is triggered, referred to as \textit{pre} filtering~\cite{craun_eliminating_2024}. The other program types reach the same granularity by filtering within the program (\textit{in}), using the \texttt{bpf\_get\_current\_cgroup\_id} helper to compare the event's \textit{cgroup} ID with a target ID; the filtering instructions then execute for every event, which may increase overhead. Filtering can also occur after the program executes (\textit{post}), collecting all events and deferring filtering to user space, serving container-granular data at query time. These methods generalize to capture approaches beyond eBPF.

\begin{figure}
\centering
\includegraphics[width=0.8\linewidth]{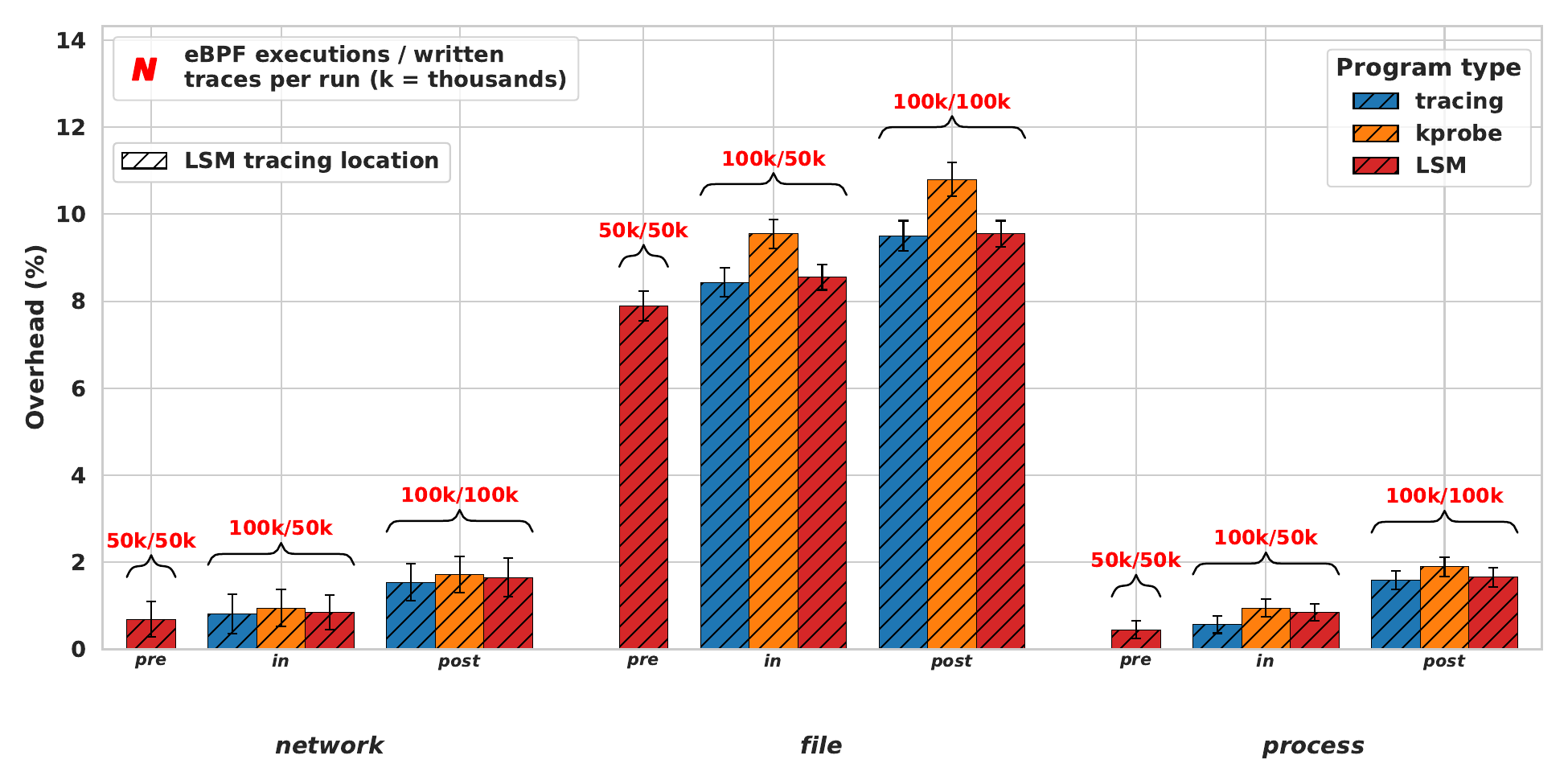}
\caption{Performance overhead as a function of the filtering used to achieve container granularity at \textit{LSM location}.}
\label{fig:experiment_filtering_location}
\end{figure}

We evaluate the overhead of these filtering methods, reusing the setup of \S~\ref{sec:experimental-setup} with one change: each container is deployed twice. Both instances run the same workload, but only one is traced; the second generates irrelevant events that \textit{in} and \textit{post} filtering must process but \textit{pre} avoids, making the overhead difference observable. Fig.~\ref{fig:experiment_filtering_location} presents the measured overhead and number of written traces per filtering method and program type (\textit{pre} applies only to \textit{LSM}, as seen before).

We observe that \textit{pre} filtering introduces the lowest overhead across all workloads. For example, under \textit{process} workload, the \textit{LSM} program incurs 0.45\% overhead with \textit{pre}, compared to 0.85\% with \textit{in} and 1.66\% with \textit{post} filtering. \textit{In} filtering is more expensive since its filtering logic executes for every event, while \textit{post} filtering yields the highest overhead: for the \textit{LSM} program, it nearly doubles the overhead of \textit{in} under \textit{network} workload (1.65\% vs.\ 0.85\%) and adds 1 percentage point under \textit{file} workload (9.55\% vs.\ 8.55\%).

Notably, \textit{LSM} programs with \textit{pre} filtering not only outperform the other program types with \textit{in} filtering, but also surpass \textit{tracing} programs, the type with the lowest per-hook cost (Fig.~\ref{fig:experiment_ebpf_program_type}). Indeed, \textit{pre} filtering achieves overheads of, respectively, 0.69\% (\textit{network}), 7.89\% (\textit{file}), and 0.45\% (\textit{process}) for the \textit{LSM} program, all lower than the corresponding values for the \textit{tracing} program with \textit{in} filtering (resp. 0.81\%, 8.43\%, and 0.57\%).
These findings highlight the efficiency advantage of direct \textit{cgroup} attachment: making \textit{LSM} programs even more attractive for \textit{provenance systems} thanks to their \textit{pre} filtering mechanism.

The red annotations report the eBPF executions and written traces per run, which explain these differences: \textit{pre} only executes for relevant events, \textit{in} executes for every event but writes only the relevant traces, and \textit{post} both executes and writes for every event, hence its higher overhead and its doubled trace volume.





\section{Classification of capture layers}
\label{sec:classification}
Building on the capture approaches and eBPF evaluation of the previous sections, we now classify existing \textit{provenance systems} and \textit{capture agents} according to these mechanisms and macro benchmark the open-source subset. We exclude \nametoolSTRACE, \nametoolALASTOR, and \nametoolARTISAN, as their user-space capture prevents any of them from satisfying all four attributes essential to security-oriented provenance simultaneously: \textit{efficiency}, \textit{visibility}, \textit{safety}, and \textit{robustness} (\S~\ref{sec:capture-approaches-taxonomy}).

\subsection{\textit{Provenance systems}}
\label{sec:provenance_systems}
Several contributions have proposed \textit{provenance systems} with distinct capture layers, each addressing a specific limitation. We observe that some capture \textit{LSM} interfaces while others capture system calls.

Among the former: CamFlow~\cite{pasquier_practical_2017,pasquier_runtime_2018} is deployed as an LSM that leverages the Netfilter and \textit{LSM} interfaces to monitor network connections alongside process and file activities. It supports fine-grained \textit{in} filtering on specific kernel objects or events. To achieve complete coverage of all information flows between kernel objects, CamFlow depends on a patched kernel that adds further \textit{LSM} interfaces; today only a subset of these proposed interfaces is in the Linux kernel.
ProvBPF~\cite{lim_secure_2021} captures at container granularity through saBPF, a patched kernel that adds kernel-object storage and lets \textit{LSM} programs attach directly to specific \textit{cgroups}, avoiding the overhead of \textit{in} or \textit{post} filtering of irrelevant host events. As discussed earlier (\S~\ref{sec:capture_granularity}), a similar capability has since been adopted in the Linux kernel 6.0~\cite{cgroupattachementbpfgitcommit}.
ConProv~\cite{deng_conprov_2024} focuses on accurately representing container activity, addressing limitations such as unreliable absolute file-path resolution within containers and the lack of security context (e.g., used capabilities). It uses \textit{tracepoint} programs to trace system calls and \textit{kprobe} programs to trace \textit{LSM} operations (e.g., \texttt{security\_capable}) tied to capability checks.

Among the systems capturing system calls:
\nametoolSPADE~\cite{gehani2012spade}, which relies on \nametoolauditd, was one of the first to model captured system calls as provenance graphs and to implement a complete capture-to-analysis pipeline. \nametoolWinnower, \nametoolTRACE, and \nametoolCLARION are all built on the \nametoolSPADE framework, though their source code has never been released. \nametoolWinnower proposes an algorithm to condense graph size without losing security-relevant semantics. \nametoolTRACE aggregates provenance capture across multiple network nodes. \nametoolCLARION targets container provenance, distinguishing mount namespaces for path names and modeling container boundaries in the graph, but does not represent orchestration-specific semantics. All of these \textit{tools} rely on \nametoolauditd, which lacks efficiency and robustness and offers only limited visibility for security provenance (\S~\ref{sec:fromthekernel} \&~\ref{sec:robustness}). eAudit~\cite{sekar_eaudit_2024} instead uses eBPF \textit{tracepoint} programs for 81 system calls (a superset of those monitored by \nametoolTRACE), mitigating data loss, overhead, and tampering through a compact encoding scheme, per-CPU ring buffers, and system-call prioritization.

\begin{figure}[h!]
\includegraphics[width=\linewidth]{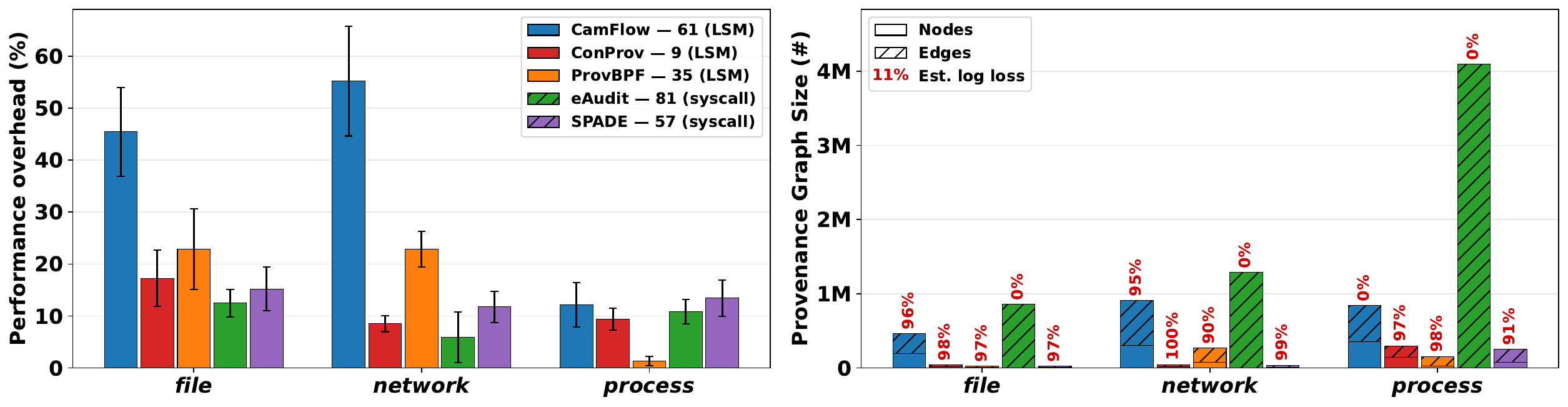}
\caption{Macro benchmark of open-source \textit{provenance systems}.}
\label{fig:experiment_provenance_capture}
\end{figure}

Beyond this survey, we empirically evaluate the open-source \textit{provenance systems}, reusing the methodology of \S~\ref{sec:experimental-setup}. Because some require a patched kernel, we deploy each on a distinct Fedora~35 server VM, the only distribution targeted by their patched kernels and build environments, which makes every tool readily reproducible. On each VM, we run the benchmark with and without capture, keeping each system's defaults. Fig.~\ref{fig:experiment_provenance_capture} reports, for every evaluated \textit{tool}, the performance overhead, captured provenance graph size, estimated log loss, and monitored kernel interfaces. We quantify log loss as $(1 - n_{\text{traced}} / n_{\text{expected}})\times 100$ over a per-workload \emph{witness event} whose ground-truth count is known (e.g., 50k \texttt{clone} events for the \textit{process} workload).

Fig.~\ref{fig:experiment_provenance_capture} shows that the evaluated systems differ widely along two axes: the performance overhead they impose and the fraction of logs they actually capture. Considering the overhead in isolation is misleading, since most systems lose a considerable share of their logs, and each one instruments a very different set of kernel interfaces (e.g., ConProv monitors only 9 kernel interfaces).

eAudit stands out: despite instrumenting the most interfaces (81 system calls), it incurs among the lowest overheads (5.9\% on \textit{network}, 12\% on \textit{file}) and loses no logs in any workload. This is because it does not reconstruct a provenance graph but emits a raw stream of system calls and their arguments; combined with per-CPU ring buffers, this keeps it efficient. Consequently, its output is an edge stream (e.g.,\ 4.1M edges in the \textit{process} workload) rather than a node/edge graph, so it never materializes the large graphs the modeling-based \textit{tools} would have produced had they not dropped logs.

The four \textit{tools} that build provenance graphs (CamFlow, ProvBPF, ConProv, and \nametoolSPADE) all suffer substantial log loss (90 to 100\% in most workloads), thus producing far smaller graphs. ConProv is the most extreme case: it instruments only 9 interfaces and loses more than 97\% of logs. \nametoolSPADE, built on \nametoolauditd, combines the highest \textit{process} overhead (13.4\%) with 91--99\% loss. CamFlow incurs the highest overhead of any evaluated system (45\% on \textit{file}, 55\% on \textit{network}) and still loses 95--96\% of logs in two workloads. It produces the largest graphs and is the only graph-modeling \textit{tool} to achieve zero log loss in the \textit{process} workload (351k nodes / 494k edges). Even with high loss, CamFlow produces large graphs due to its modeling choices: it represents files, inodes, and paths as distinct graph nodes, so even a 96\% log loss yields a big \textit{file} graph (192k nodes / 272k edges); similarly, modeling every network packet as its own node produces the largest \textit{network} graph (300k nodes / 610k edges) despite roughly 95\% of logs being lost.

Overall, no evaluated \textit{provenance system} reconciles overhead, log loss, and actually modeling the graph: the only \textit{system} that avoids log loss and is efficient (i.e., eAudit) omits graph construction.

\subsection{\textit{Capture agents}}
\label{sec:capture_agent}
Beyond the \textit{provenance systems} above, several widely adopted \textit{capture agents} collect kernel telemetry for broader observability. Unlike \textit{provenance systems}, which build provenance graphs over long time windows, these agents monitor for limited windows (e.g., debugging or incident response) and can thus afford to trace broader event sets, up to every system call. Serving a different purpose, they cannot be directly compared to \textit{provenance systems}; we therefore place their macro benchmark in Appendix~\ref{sec:appendix_macrobench_capture_agents}. In sum, monitoring a large, system-call-heavy interface set yields large log volumes and considerable log loss.

LTTng~\cite{noauthor_lttng_2005} is an out-of-tree kernel module capturing system calls and kernel events via \textit{tracepoints}, kernel functions via \textit{kprobes}, and user-space telemetry from Python, Java, and C/C++; unlike eBPF, such modules offer weaker safety, an acceptable trade-off given LTTng's maturity. Sysdig, Tracee, and Tetragon are eBPF-based agents for granular observability and rule-based detection. Sysdig traces system calls with \textit{tracing} programs and other kernel events (e.g., page faults) with \textit{tracing} and \textit{tracepoint}, for debugging and security monitoring. Tracee ships predefined \textit{LSM}, \textit{kprobe}, \textit{tracepoint}, and \textit{tracing} programs for specific security, network, and system events. Tetragon targets SOC teams and administrators with an operator-friendly YAML DSL specifying and enforcing capture policies, compiled into \textit{LSM}, \textit{kprobe}, \textit{tracepoint}, and \textit{uprobe} programs.

Their strengths for provenance are
\begin{inparaenum}
\item a mature capture model with predefined programs identifying high-level system events (e.g., container creation or kernel module loading), and
\item strong industrial support ensuring portability.
\end{inparaenum}

\begin{table}[h!]
\caption{\textit{Provenance systems}.}
\label{tab:comparison_complete_provenance_tools_container}
\resizebox{\textwidth}{!}{
  \begin{tabular}{cccccccc@{\hspace{1em}}c}
    \toprule
    \makecell{\textbf{Provenance}\\\textbf{system}} & \makecell{\textbf{Capture}\\\textbf{approach}} & \multicolumn{2}{c}{\makecell{\textbf{Cloud-native}\\\textbf{awareness}}}& \multicolumn{3}{c}{\textbf{Capture granularity}}&\makecell{\textbf{Open source}\\\textbf{availability}} & \textbf{Maintained} \\
    &&container&orchestration&\textit{pre}&\textit{in}&\textit{post}&&\\
    \midrule
    \makecell{\nametoolSPADE~\cite{gehani2012spade}\\(2012)}&\nametoolauditd&\xmark&\xmark&system call&$\emptyset$&event attribute&\cmark&\cmark\\
    \midrule
    \makecell{CamFlow~\cite{pasquier_practical_2017}\\(2017)}&\makecell{LSM\\(patched \textit{LSM}, Netfilter)}&\xmark&\xmark&\makecell{MAC operation\\network connection}&\makecell{kernel object\\kernel event}&\(\emptyset\)&\cmark&\xmark\\
    \midrule
    \makecell{\nametoolWinnower~\cite{hassan_towards_2018}\\(2018)}&\nametoolauditd&\cmark&\xmark&system call&$\emptyset$&$\emptyset$&\xmark&\xmark\\
    \midrule
    \makecell{\nametoolTRACE~\cite{irshad_trace_2021}\\(2021)}&\nametoolauditd&\xmark&\xmark&system call&$\emptyset$&$\emptyset$&\xmark&\xmark\\
    \midrule
    \makecell{ProvBPF~\cite{lim_secure_2021}\\(2021)}&\makecell{eBPF\\(patched \textit{LSM})}&\midsym&\xmark&\makecell{MAC operation\\container}&\(\emptyset\)&\(\emptyset\)&\cmark&\xmark\\
    \midrule
    \makecell{\nametoolCLARION~\cite{chen_clarion_2021}\\(2021)}&\nametoolauditd&\cmark&\xmark&system call&container&\(\emptyset\)&\xmark &\xmark\\
    \midrule
    \makecell{ConProv~\cite{deng_conprov_2024}\\(2024)}&\makecell{eBPF\\(\textit{kprobe}, \textit{tracepoint})}&\cmark&\xmark&\makecell{system call\\capabilities\\socket connection\\file permission}&container&\(\emptyset\)&\cmark&\xmark\\
    \midrule
    \makecell{eAudit~\cite{sekar_eaudit_2024}\\(2024)}&\makecell{eBPF\\(\textit{tracepoint})}&\xmark&\xmark&system call&\(\emptyset\)&\(\emptyset\)&\cmark&\xmark\\
    \bottomrule
\end{tabular}}
\end{table}

\begin{table}[h!]
\caption{\textit{Capture agents} that could serve the capture layer of \textit{provenance systems}.}
\label{tab:comparison_capture_agents}
\resizebox{\textwidth}{!}{
  \begin{tabular}{cccccccc@{\hspace{1em}}c}
    \toprule
    \makecell{\textbf{Capture}\\\textbf{agent}} & \makecell{\textbf{Capture}\\\textbf{approach}} &\multicolumn{2}{c}{\makecell{\textbf{Cloud-native}\\\textbf{awareness}}}& \multicolumn{3}{c}{\textbf{Capture granularity}}& \makecell{\textbf{Open source}\\\textbf{availability}}&\textbf{Maintained}\\
    &&container&orchestration&\textit{pre}&\textit{in}&\textit{post}&& \\
    \midrule
    \makecell{\nametoolauditd~\cite{noauthor_auditd_2004}\\(2004)}&\makecell{kernel audit\\subsystem\\(\textit{syscall}, \textit{LSM})}&\xmark&\xmark&\makecell{system call\\MAC operation\\file watch}&\(\emptyset\)&event attribute&\cmark&\cmark\\
    \midrule
    \makecell{LTTng~\cite{noauthor_lttng_2005}\\(2005)}&\makecell{out-of-tree kernel module\\(\textit{kprobe}, \textit{tracepoint},\\programming libraries)}&\cmark&\xmark&\makecell{system call\\kernel function\\kernel event\\C/C++, Java \&\\Python program}&\(\emptyset\)&\makecell{event attribute\\
    (incl. container)}&\cmark&\cmark\\
    \midrule
    \makecell{Sysdig~\cite{noauthor_sysdig_2014}\\(2014)}&\makecell{eBPF\\(\textit{tracing, tracepoint})}&\cmark&\xmark&\makecell{system call\\kernel event}&\(\emptyset\)&\makecell{event attribute\\(incl. container)}&\cmark&\cmark\\
    \midrule
    \makecell{Tracee~\cite{noauthor_tracee_2020}\\(2020)}&\makecell{eBPF\\(\textit{LSM}, \textit{tracing},\\\textit{kprobe}, \textit{tracepoint})}&\cmark&\cmark&\makecell{MAC operation\\system call\\kernel event\\network event\\}&\makecell{process, binary\\namespace\\container, pod\\\textit{cgroup}, user\\}&\(\emptyset\)&\cmark&\cmark\\
    \midrule
    \makecell{Tetragon~\cite{noauthor_tetragon_2022}\\(2022)}&\makecell{eBPF\\(\textit{LSM}, \textit{kprobe},\\\textit{tracepoint}, \textit{uprobe})}&\cmark&\cmark&\makecell{MAC operation\\system call\\kernel function\\user-space function}&\makecell{func. argument\\process, binary\\namespace\\container}&event attribute&\cmark&\cmark\\
    \bottomrule
\end{tabular}}
\end{table}

\subsection{Classification}
\label{sec:classification-results}
To compare these \textit{provenance systems} and \textit{capture agents}, we classify them in Tables~\ref{tab:comparison_complete_provenance_tools_container} and~\ref{tab:comparison_capture_agents} along five key attributes of their capture layer: capture approach, cloud-native awareness, capture granularity, open-source availability, and maintenance. We discuss each one by drawing on the previous empirical evaluations.

\subsubsection{Capture approach}(Column~2): mechanism to capture kernel telemetry.

Capture mostly occurs in the kernel via eBPF, dominant among the most recent contributions, in line with the benefits analyzed in \S~\ref{sec:fromthekernel}. Several \textit{provenance systems}, however, still build on \nametoolauditd (i.e., \nametoolSPADE, \nametoolWinnower, \nametoolTRACE, and \nametoolCLARION among those surveyed), despite the limitations discussed above (\S~\ref{sec:fromthekernel} \&~\ref{sec:robustness}). Within these eBPF-based \textit{tools}, capture mostly relies on the \textit{LSM}, \textit{tracepoint}, or \textit{kprobe} program types (\S~\ref{sec:fromthekernel}), with \textit{provenance systems} attaching to both entry and exit points of system calls and sometimes adding \textit{kprobe} interfaces for internal kernel functions. The more recent \textit{tracing} program type remains used almost exclusively by \textit{capture agents}, even though its trampoline-based attachment offers the lowest per-attachment overhead we measured (\S~\ref{sec:overhead-program-type}).

Beyond the capture approach, the set of instrumented interfaces differs by \textit{tool} class: \textit{provenance systems} select tailored subsets, whereas \textit{capture agents} instrument broader sets for visibility. Effective provenance, however, requires capturing not only system calls but also kernel-object life cycles (e.g., inode allocation) and information flows between objects~\cite{pasquier_practical_2017}. \textit{LSM} interfaces naturally cover these life cycles and most information flows, but their completeness remains an open question: Georget et al.~\cite{georget_verifying_2017,georget_kayrebt_2015} used compile-time control-flow graphs to verify which system-call paths reach \textit{LSM} interfaces and, on Linux~4.3, found missing interfaces for pipe, splice, repeated socket messages, memory unmapping, and shared-memory detachments, a subset of which are now in the kernel.

In sum, we observe that \textit{LSM} interfaces are the most attractive basis for capture: they offer low overhead (\S~\ref{sec:ebpf-empirical-evaluation} \&~\ref{sec:capture_granularity}), expose kernel-object insights, can be extended for missing information flows, and their smaller surface (265 \textit{LSM} vs.\ 407 system calls in Linux~6.14, counting every interface regardless of configuration or architecture) yields more abstract graphs with fewer events while preserving semantic richness. System-call-based systems, by contrast, diverge widely in the set they capture, and none assess whether their chosen subset suffices for complete provenance graphs.


\subsubsection{Cloud-native awareness}(Column~3) assesses whether a \textit{provenance system} or \textit{capture agent} records container- and orchestration-specific events. Container-aware \textit{tools} should trace container runtime entities and their attributes (e.g., IDs, labels, resource constraints, life-cycle states), while orchestration-aware \textit{tools} should additionally capture orchestration-specific events such as pod life-cycles.

\textit{Capture agents} are generally container-aware, the result of the significant engineering effort behind these large, industry-backed open-source projects; \nametoolauditd is the exception, a long-standing kernel audit facility that predates containers and records neither container nor orchestration semantics. \textit{Provenance systems} lag behind. As Deng et al.~\cite{deng_conprov_2024} note, many of them, such as ProvBPF, identify containers as entities but lack complete records of container-specific security metadata, such as the use of Linux capabilities. Container-awareness is nonetheless attainable with \nametoolauditd: \nametoolWinnower and \nametoolCLARION group audit events per container and are able to distinguish them from the host.
Orchestration-awareness remains exclusive to Tracee and Tetragon.

\subsubsection{Capture granularity}(Column~4) evaluates how far a system filters events to capture only relevant subsets of provenance data. Granularity can apply to objects (e.g., processes, users, or containers), events (e.g., system calls or scheduler activity), or event attributes (e.g., timestamps), filtered before capture (\textit{pre}), during instrumentation (\textit{in}), or after capture (\textit{post}); \(\emptyset\) indicates no filtering.

Event granularity is mostly achieved through \textit{pre} filtering, by attaching capture only to the relevant interfaces. Container granularity, by contrast, mostly relies on the costlier \textit{in} or \textit{post} filtering; the notable exception is ProvBPF, whose patched saBPF kernel attaches eBPF programs directly to a \textit{cgroup}, reaching container granularity through \textit{pre} filtering.

\subsubsection{Open source availability \& maintenance}(Columns~5 \& 6) examine whether each \textit{tool}'s source code is public and if it is actively maintained after publication.

In contrast to the \textit{capture agents}, some \textit{provenance systems} are not open source, and all except \nametoolSPADE are unmaintained. Even among the open-source ones, portability is limited: CamFlow and ProvBPF depend on custom kernel patches, restricting their deployment to a specific kernel compiled with specific modules and configurations. \textit{Capture agents}, by contrast, are actively developed and frequently commercially backed, offering the stability and long-term usability that make them more suitable foundations for \textit{provenance systems}.

\section{Conclusion}
We studied the kernel telemetry options for building the capture layer of security-oriented \textit{provenance systems}, working from the capture mechanisms up to the \textit{tools} that build on them: we classified five capture approaches, empirically benchmarked the eBPF program types and the filtering methods that achieve capture granularity, and classified eight \textit{provenance systems} and five \textit{capture agents}, running our macro benchmark on those available as open source.

On \emph{how} to capture, eBPF is now the de facto mechanism, but its program types differ in cost, and \textit{LSM} programs emerge as the best suited for provenance: they capture a MAC operation's arguments and verdict with a single attachment, where every other program type must hook both entry and exit, whether of the system call or of the \texttt{security\_} wrapper, expose stable interfaces and reach \textit{pre} filtering through direct \textit{cgroup} attachment. Capturing the verdict is not optional for the use case: an operation recorded at entry may still be denied by an LSM such as SELinux, and while these blocked attempts inform observability, a provenance graph must record only the causal events that actually occurred.

On \emph{what} to capture and on \emph{robustness}, the most overlooked dimension, \textit{tools} fall short. They instrument broader interface sets than the use case requires, yet none verify that their subset captures provenance completely; \textit{LSM} interfaces are nonetheless an attractive basis, exposing kernel-object life cycles over a smaller surface. On robustness, neither integrity nor availability is guaranteed: system-call-based \textit{tools} remain open to TOCTOU evasion, and every graph-building \textit{tool} drops most of its logs in our macro benchmark. \textit{LSM} programs, which inspect arguments already copied into the kernel, offer the strongest integrity guarantees.

Overall, \textit{provenance systems} rely on widely different capture layers, most unable to guarantee event integrity and availability, leaving them unsuitable for security use cases; \textit{capture agents}, by contrast, offer the maintenance and portability that \textit{provenance systems} lack, making them attractive building blocks for future systems, though most remain slow and lossy. A fair evaluation must ultimately look beyond capture to each \textit{tool}'s configurability and the expressiveness of its provenance output: future work should assess the semantic quality of the resulting graphs for downstream security tasks through metrics such as intrusion-detection accuracy and forensic completeness.

\bibliographystyle{plainnat}
\bibliography{references}

\newpage
\appendix
\section{Robustness analysis}
\label{sec:appendix_robustness}
\begin{table}[h!]
\centering
\begin{subtable}[t]{0.6\linewidth}
  \centering
  \tiny
  \setlength{\tabcolsep}{3pt}
  \begin{tabular}{>{\raggedright\arraybackslash}p{0.62\linewidth}>{\raggedright\arraybackslash}p{0.32\linewidth}}
    \toprule
    Syscalls & User-space pointer(s) \\
    \midrule
    \texttt{accept}, \texttt{accept4} & \texttt{upeer\_sockaddr}, \texttt{upeer\_addrlen} \\
    \midrule
    \texttt{bpf} & \texttt{attr} \\
    \midrule
    \texttt{chmod}, \texttt{fchmodat}, \texttt{open}, \texttt{openat} & \texttt{filename} \\
    \midrule
    \texttt{connect} & \texttt{uservaddr} \\
    \midrule
    \texttt{creat}, \texttt{mkdir}, \texttt{mkdirat}, \texttt{rmdir}, \texttt{unlink}, \texttt{unlinkat} & \texttt{pathname} \\
    \midrule
    \texttt{dup}, \texttt{dup2}, \texttt{dup3}, \texttt{fchmod}, \texttt{listen}, \texttt{ptrace}, \texttt{setns}, \texttt{setuid}, \texttt{socket}, \texttt{unshare}, \texttt{userfaultfd} & $\emptyset$ \\
    \midrule
    \texttt{execve}, \texttt{execveat} & \texttt{filename}, \texttt{argv}, \texttt{envp} \\
    \midrule
    \texttt{finit\_module} & \texttt{uargs} \\
    \midrule
    \texttt{init\_module} & \texttt{umod}, \texttt{uargs} \\
    \midrule
    \texttt{link}, \texttt{linkat}, \texttt{rename}, \texttt{renameat}, \texttt{renameat2}, \texttt{symlinkat} & \texttt{oldname}, \texttt{newname} \\
    \midrule
    \texttt{openat2} & \texttt{filename}, \texttt{how} \\
    \midrule
    \texttt{recvfrom} & \texttt{ubuf}, \texttt{addr}, \texttt{addr\_len} \\
    \midrule
    \texttt{recvmsg}, \texttt{sendmsg} & \texttt{msg} \\
    \midrule
    \texttt{sendto} & \texttt{buff}, \texttt{addr} \\
    \midrule
    \texttt{symlink} & \texttt{old}, \texttt{new} \\
    \bottomrule
  \end{tabular}
  \caption{Falco}
\end{subtable}
\hfill
\begin{subtable}[t]{0.39\linewidth}
  \centering
  \tiny
  \setlength{\tabcolsep}{3pt}
  \begin{tabular}{>{\raggedright\arraybackslash}p{0.52\linewidth}>{\raggedright\arraybackslash}p{0.42\linewidth}}
    \toprule
    Syscalls & User-space pointer(s) \\
    \midrule
    \texttt{connect} & \texttt{uservaddr} \\
    \midrule
    \texttt{dup}, \texttt{dup2}, \texttt{dup3}, \texttt{mprotect}, \texttt{ptrace} & $\emptyset$ \\
    \midrule
    \texttt{execve}, \texttt{execveat} & \texttt{filename}, \texttt{argv}, \texttt{envp} \\
    \midrule
    \texttt{finit\_module} & \texttt{uargs} \\
    \midrule
    \texttt{init\_module} & \texttt{umod}, \texttt{uargs} \\
    \midrule
    \texttt{mount} & \texttt{dev\_name}, \texttt{dir\_name}, \texttt{type}, \texttt{data} \\
    \midrule
    \texttt{open}, \texttt{openat} & \texttt{filename} \\
    \midrule
    \texttt{open\_by\_handle\_at} & \texttt{handle} \\
    \midrule
    \texttt{openat2} & \texttt{filename}, \texttt{how} \\
    \midrule
    \texttt{rename}, \texttt{renameat2} & \texttt{oldname}, \texttt{newname} \\
    \midrule
    \texttt{write} & \texttt{buf} \\
    \midrule
    \texttt{writev} & \texttt{vec} \\
    \bottomrule
  \end{tabular}
  \caption{Tracee}
\end{subtable}
\caption{System calls referenced by the Falco and Tracee attack signatures and their user-space (\texttt{\_\_user}) pointer arguments.}
\label{tab:robustness}
\end{table}

\FloatBarrier
\section{Macro benchmark of open-source \textit{capture agents}}
\label{sec:appendix_macrobench_capture_agents}

\begin{figure}[h!]
\centering
\includegraphics[width=\linewidth]{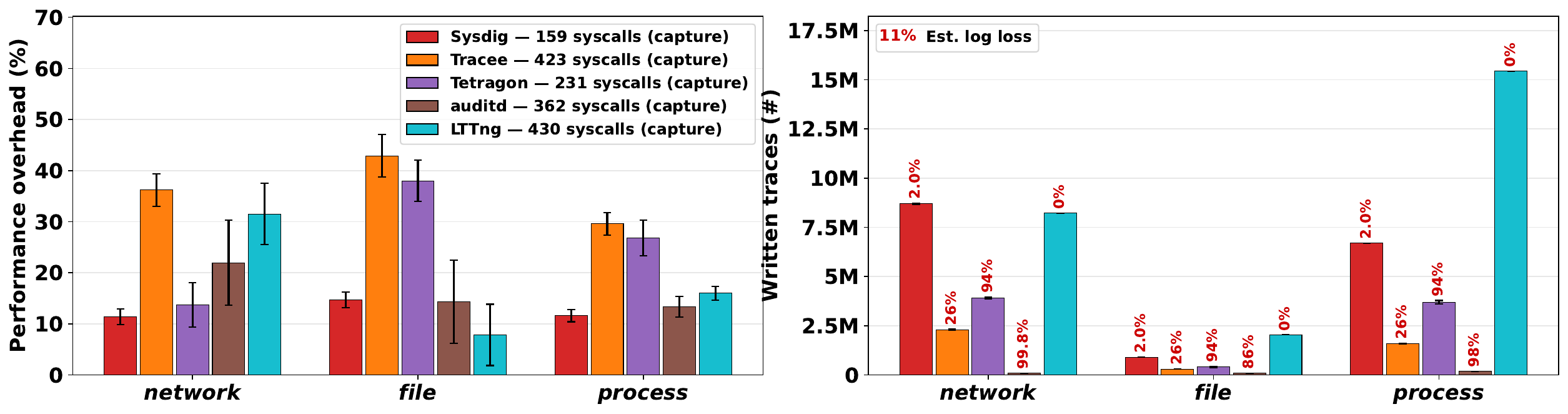}
\caption{Macro benchmark of open-source \textit{capture agents}.}
\label{fig:macro_bench_capture}
\end{figure}

\end{document}